\documentclass[a4paper,11pt]{article}
\usepackage{amsmath, amsfonts, amssymb, comment, amsthm}

\title{Isometry and Automorphisms of Constant Dimension Codes}
\author{Anna-Lena Trautmann \\ Institute of Mathematics\\ University of Zurich}
\date{}

\newtheorem{thm}{Theorem}
\newtheorem{cor}[thm]{Corollary}
\newtheorem{prop}[thm]{Proposition}
\newtheorem{lem}[thm]{Lemma}

\theoremstyle{definition}
\newtheorem{defi}[thm]{Definition}

\newtheorem{ex}[thm]{Example}

\newcommand{\rs}{\mathrm{rs}}
\newcommand{\F}{\mathbb{F}}
\newcommand{\G}{\mathcal{G}_{q}(k,n)}

\newcommand{\ProG}{\mathrm{PG}(q,n)}
\newcommand{\U}{\mathcal{U}}
\newcommand{\V}{\mathcal{V}}
\newcommand{\C}{\mathcal{C}}
\newcommand{\Aut}{\mathrm{Aut}}
\newcommand{\SAut}{\mathrm{SAut}}
\newcommand{\dbl}{/ \!\! /}

\newcommand{\Stab}[2]{\mathrm{Stab}_{{#1}}\left({#2}\right)}%
\newcommand{\Mat}[2]{\mathrm{Mat}_{#1 \times #2}}
\newcommand{\GL}[1]{\mathrm{GL}_{#1}}
\newcommand{\GammaL}[1]{\mathrm{\Gamma L}_{#1}}
\newcommand{\PGL}[1]{\mathrm{PGL}_{#1}}
\newcommand{\PGammaL}[1]{\mathrm{P\Gamma L}_{#1}}

\begin{document}

\maketitle

\begin{abstract}
We define linear and semilinear isometry for general subspace codes, used for random network coding. Furthermore, some results on isometry classes and automorphism groups of known constant dimension code constructions are derived. 
\end{abstract}

\section{Introduction}

\emph{Subspace codes} are used for random linear network coding  \cite{ah00, ko08}. They are defined as subsets of the projective geometry, which is the set of all subspaces of a given ambient space over a finite field.
In the special case that all codewords have the same dimension, we call those codes \emph{constant dimension codes}. It makes sense to define isometry classes of these codes and a canonical representative of each class to compare codes among each other.

On the other hand, a canonical form and the automorphism group are important for the theory of orbit codes \cite{tr10p},  which are a special family of constant dimension codes. These codes are defined as orbits of a subgroup of the general linear group on an element of the projective geometry over a finite field. Different subgroups can possibly generate the same orbit, hence one needs a canonical way to compare orbit codes among each other. This can be done via the automorphism groups of the codes, since these are the maximal generating groups for a given orbit code and they contain all other generating subgroups of it.

The paper is structured as follows: We give some preliminary results in Section \ref{sec:prelim}, first on network coding in general and on orbit codes. The second part of the section deals with linear and semilinear isometry of general subspace codes. It is shown that any (semi-)linearly isometric code of a given code can be reached by action of the projective (semi-)linear group. 

In Section \ref{sec:isoex} we derive some theoretical results on and give some examples of isometry classes and automorphism groups of spread codes, orbit codes and lifted rank metric codes, which are known code constructions that will be explained in detail in that part.

We conclude in Section \ref{sec:conclusion} by summing up the results.


\section{Preliminaries}\label{sec:prelim}

\subsection{Network Coding}

Let $\mathbb{F}_q$ be the finite field with $q$ elements and the projective geometry $\ProG$ the set of all subspaces of $\F_{q}^{n}$, whereas $\G$ is the set of all
subspaces of $\mathbb{F}_q^n$ of dimension $k$, called
\emph{Grassmannian}. The general linear group $\GL{n}(q)$ is the set of all invertible $n\times n$-matrices with entries in $\F_{q}$. 

$\Aut(\F_q)$ denotes the automorphism group of $\F_q$. Recall that any automorphism $\alpha$ of a finite field of characteristic $p$ is of the type $\alpha(x) = x^{p^j}$. It applies to vectors and matrices element-wise. Denote by $Gal(\F_{q^k}, \F_q)$ the Galois group of $\F_{q^k}$ over $\F_q$, i.e. the set of all automorphisms of $\F_{q^k}$ that stabilize the subfield $\F_q$. If $p$ is the characteristic of $\F_q$, then it holds that $\Aut(\F_q)=Gal(\F_q, \F_p)$.

The set of all semilinear mappings, i.e. the general semilinear group $\GammaL{n}(q) := \GL{n}(q)\rtimes \Aut(\F_q)$ decomposes as a semidirect product with the multiplication
\[(A,\alpha) (B, \beta) := (A \: \alpha^{-1} (B), \alpha \beta).\]

By $\Mat{k}{n}(q)$ we denote the set of all $k\times n$-matrices with entries in $\F_{q}$. If the underlying field is clear from the context we abbreviate the above by $\GL{n}, \GammaL{n}$ and $\Mat{k}{n}$, respectively.

Let $U\in \Mat{k}{n}$ be a matrix of rank $k$ and
\[\mathcal{U}=\rs (U):= \text{row space}(U)\in \G.\] 
One notices that the row space is invariant under $\GL{k}$-multiplication on the left, i.e. for any $T\in \GL{k}$
\[\mathcal{U}=\rs(U)= \rs(T U).\] 
A unique representative of all matrices with the same row space is the one in reduced row echelon form.
Any $k\times n$-matrix can be transformed into reduced row echelon form by a unique $T\in \GL{k}$.

The \emph{subspace distance} and the \emph{injection distance} are metrics on the projective geometry $\ProG$ given by
\begin{align*}
d_S(\mathcal{U},\mathcal{V}) =& \dim(\U + \V) - \dim(\U \cap \V)\\
=& \dim(\U) + \dim(\V) - 2\dim(\mathcal{U}\cap
\mathcal{V})
\end{align*}
\begin{align*}
d_I(\mathcal{U},\mathcal{V}) =&\max \{\dim(\U) ,\dim(\V)\} - \dim(\mathcal{U}\cap
\mathcal{V})
\end{align*}
for any $\mathcal{U},\mathcal{V} \in \ProG$. They are suitable distances for coding over the operator channel \cite{ko08}, where the injection metric is the more suitable one for an adversary model \cite{si09}. Since for $\U, \V \in \G$ it holds that
\[d_S(\U,\V) = 2 d_I(\U,\V) ,\]
they are exchangeable in the study of constant dimension codes. If we do not need to specify which metric we are using we will write $d(\U,\V)$.

In general a  \emph{subspace code} is simply a subset of $\ProG$. 
A \emph{constant dimension code} is a subset of $\G$. The minimum distance of a code is defined in the usual way. 

Bounds on the size of subspace codes can be found in \cite{et08p, kh09, ko08}. Different constructions of constant dimension codes have been investigated in e.g. \cite{et08u, ko08p, ko08, si08a, sk10, tr10p}.

Given $U\in \Mat{k}{n}$ of rank $k$,
$\mathcal{U}\in \G$ its row space and $(A, \alpha) \in
\GammaL{n}$, we define
\[\mathcal{U} (A,\alpha):= \rs(\alpha(U A)).\] 
Since $\alpha(TU A) = \alpha(T) \alpha(U A)$ for any $T\in \GL{k}$, the operation here defined is independent from the representation of $\mathcal{U}$ and therefore well-defined. The $\GammaL{n}$-multiplication defines a group action from the right on the Grassmannian and hence on $\ProG$ as well:
\[
\begin{array}{ccc}
\G\times \GammaL{n}& \longrightarrow &
\G \\ 
(\mathcal{U}, (A,\alpha)) & \longmapsto & \mathcal{U} (A,\alpha)
\end{array}
\]
This is indeed a group action since
\[(\U (A, \alpha))(B,\beta) = (\alpha(\U A))(B,\beta) = \beta (\alpha(\U A) B) =\beta  \alpha ((\U A) \alpha^{-1}(B)) \]\[= \alpha\beta  (\U (A \alpha^{-1}(B)))= \U (A\alpha^{-1}(B), \alpha \beta) = \U ((A,\alpha)(B,\beta))
.\]
It induces a group action of $\GL{n}$ on $\ProG$, too. 

This action respects the distances $d_S, d_I$ and therefore defines a notion of equivalence for subspace codes. In Section \ref{subsec:isometry} we will show, that this equivalence is the most general one may demand if one also wants to preserve some other elementary properties of random subspace codes.

Generally, group actions on sets are performed element-wise. For a group $G$ acting from the right on a set $X$ and an element $x\in X$, $\Stab{G}{x}:=\{g\in G \mid xg = x\}$ denotes the stabilizer of $x$ under $G$. The orbit of $G$ on $x\in X$ is denoted by $xG:=\{xg\mid g\in G\}$ and the set of all orbits by $X\dbl G:=\{x G\mid x \in X\}$. A transversal of $X\dbl G$ is a set containing one element of each orbit.


An orbit $\U G, G \leq \GL{n}$ on a point $\U$ of the Grassmannian is also called an \emph{orbit code} \cite{tr10p}. Since
\[\G \cong \GL{n}/\Stab{\GL{n}}{\mathcal{U}} \]
it is possible that different groups generate the same orbit code. 

For the whole paper we will use vectors in row form and, if not stated differently, $\GammaL{n}$ and $\GL{n}$ will be applied from the right.


\subsection{Isometry of Subspace Codes} \label{subsec:isometry}

An open question is how to define equivalence of subspace codes. Naturally equivalent codes should have the same ambient space, cardinality, error-correction capability (i.e. minimum distance) and transmission rate (for a fixed ambient space this is given by the  maximal dimension of the codewords). Moreover, the distance distribution and the dimension distribution should be the same. Clearly, these last two conditions imply the minimum distance and maximum dimension.

This work engages in the isomorphic (with respect to the subset relation) equivalences of subspace codes.

\begin{defi}
A distance-preserving map
$
\iota : \ProG \rightarrow \ProG
$
i.e. fulfilling
\begin{align*}
d(\mathcal{U},\mathcal{V}) = d(\iota(\mathcal{U}),\iota(\mathcal{V})) \quad \forall \; \mathcal{U},\mathcal{V} \in \ProG.
\end{align*}
is called an \emph{isometry} on $\ProG$. 
\end{defi}

Any isometry $\iota$ is injective:
\begin{align*}
\mathcal{U} \ne \mathcal{V} \Longleftrightarrow d(\mathcal{U},\mathcal{V}) \ne 0 
\Longleftrightarrow d(\iota(\mathcal{U}),\iota(\mathcal{V})) \ne 0 \Longleftrightarrow \iota(\mathcal{U}) \ne \iota(\mathcal{V})
\end{align*}
and hence, if the domain is equal to the co-domain, bijective. The inverse map $\iota^{-1}$ is an isometry as well.

\begin{lem}
If $\iota : \ProG \rightarrow \ProG$ is an isometry, then $\iota( \{0\}) \in \left\{ \{0\}, \F_{q}^{n} \right\}$.
\end{lem}

\begin{proof}
We will prove it using the subspace distance. The proof for the injection metric is analogous.

Assume $ \mathcal{U}:= \iota( \{0\})  \not \in \left\{ \{0\}, \F_{q}^{n}\right\}$ and let $\mathcal{V} := \iota(\F_{q}^{n})$. It holds that
\begin{align*}
d_S( \{0\}, \F_{q}^{n}) &= d_{S}(\iota (\{0\}), \iota(\F_{q}^{n})) \\ \Longleftrightarrow \hspace{2cm}
n &= d_S(\mathcal{U}, \mathcal{V}) \\  \Longleftrightarrow \hspace{2cm}
n &= \dim(\mathcal{U} + \mathcal{V}) - \dim(\mathcal{U} \cap \mathcal{V}) .
\end{align*}
This implies $\mathcal{U} + \mathcal{V} = \F_{q}^{n}$ and $\mathcal{U} \cap \mathcal{V} = \{0\}$ and thus $\mathcal{V} \not \in \left\{ \{0\}, \F_{q}^{n}\right\}$. Choose non-zero vectors $u \in \mathcal{U}, v\in\mathcal{V}$ and consider the one-dimensional subspace $\mathcal{W}$ generated by $u+v$. Then $\dim (\U \cap \mathcal{W}) = \dim (\V \cap \mathcal{W})=0$ and
\begin{align*}
d_S( \iota^{-1}(\mathcal{W}),\{0\}) = d_S(\mathcal{W}, \mathcal{U}) = 1+\dim(\mathcal{U}) \\
d_S( \iota^{-1}(\mathcal{W}), \F_{q}^{n}) = d_S(\mathcal{W}, \mathcal{V}) = 1+\dim(\mathcal{V})
\end{align*}
which leads to the following contradiction (recall that $d_S(\mathcal{X},\{0\})= \dim(\mathcal{X})$ and $d_S(\mathcal{X},\F_{q}^{n})=  n-\dim(\mathcal{X})$ for any $\mathcal{X} \in \ProG$):
\begin{align*}
n =& d_S( \iota^{-1}(\mathcal{W}),\{0\}) + d_S( \iota^{-1}(\mathcal{W}), \F_{q}^{n}) 
= 2+\dim(\mathcal{U})+ \dim(\mathcal{V}) = 2+n
\end{align*}
\end{proof}

\begin{lem}\label{lem3}
Let $\iota$ be as before and $\U \in \ProG$ arbitrary. Then
\[\iota( \{0\})= \{0\}  \Longrightarrow  \dim(\U) = d(\{0\}, \U) = d(\{0\}, \iota(\U))  =  \dim (\iota(\U))  \]
and on the other hand
\[\iota( \{0\})= \F_{q}^{n}  \Longrightarrow  \dim(\U) = d(\{0\}, \U) = d(\F_{q}^{n}, \iota(\U))  = n-\dim (\iota(\U))  .\]
\end{lem}

In the following, we restrict ourselves to the isometries with $\iota( \{0\})= \{0\}$ because these are exactly the isometries that keep the dimension of a codeword. Now we want to characterize all isometries on $\ProG$ with $\iota( \{0\})= \{0\}$. 
For it we need the Fundamental Theorem of Projective Geometry (cf. \cite{ar88b,ba52b}): 

\begin{thm}\label{ftpg}
Let $\mathcal{Z}_n := \{ \mu I_n \mid \mu \in \F_q^{\ast}\}$ be the set of scalar transformations. Then every order-preserving bijection (with respect to the subset relation) $f: \ProG \rightarrow \ProG$, where $n > 2$, is induced by a semilinear transformation $(A, \alpha)\in$
\[ \mathrm{P}\Gamma \mathrm{L}_{n} = \left( \GL{n} / \mathcal{Z}_{n} \right) \rtimes \Aut(\F_q) .
\]
\end{thm}

\begin{thm}
For $n> 2$ a map $\iota  : \ProG \rightarrow \ProG$ is an order-preserving bijection (with respect to the subset relation) of $\ProG$ if and only if it is an isometry  with $\iota(\{0\}) = \{0\}$. 
\end{thm}

\begin{proof} 
We will again prove the statement using the subspace distance, where an analogous proof holds for the injection distance.
\begin{enumerate}
\item ``$\Longleftarrow$'' \\
Let $\iota$ be an isometry with $\iota(\{0\}) = \{0\}$.
We have to show that for any $\mathcal{U}, \mathcal{V} \in \ProG$ the following holds:
\begin{align*}
\mathcal{U} \subseteq \mathcal{V} \Longleftrightarrow \iota(\mathcal{U}) \subseteq \iota(\mathcal{V})
\end{align*}
From Lemma \ref{lem3} one knows that $\dim(\mathcal{U}) = \dim(\iota(\mathcal{U}))$.
Assume that there are $\mathcal{U}, \mathcal{V} \in \ProG$ with $\mathcal{U} \subseteq \mathcal{V}$ and $\iota(\mathcal{U}) \not\subseteq \iota(\mathcal{V})$.
This leads to the contradiction:
\begin{align*}
d_S( \iota(\mathcal{U}), \iota(\mathcal{V}) ) &= \dim(\iota(\mathcal{U})) + \dim(\iota(\mathcal{V})) - 2\dim(\iota(\mathcal{U}) \cap \iota(\mathcal{V}))\\ 
& > \dim(\iota(\mathcal{U})) + \dim(\iota(\mathcal{V})) - 2\dim(\iota(\mathcal{U})) \\
& = \dim(\mathcal{U}) + \dim(\mathcal{V}) - 2\dim(\mathcal{U})\\
& = \dim(\mathcal{U}) + \dim(\mathcal{V}) - 2\dim(\mathcal{U}\cap \V)\\
& = d_S( \mathcal{U}, \mathcal{V})
\end{align*}

Hence $
\mathcal{U} \subseteq \mathcal{V}\Longrightarrow \iota(\mathcal{U}) \subseteq \iota(\mathcal{V}) $. Since $\iota^{-1}$ is an isometry as well, the converse %
also holds. Thus, $\iota $ is an order-preserving bijection.

\item ``$\Longrightarrow$'' \\
According to Theorem \ref{ftpg} any order-preserving bijection $\iota$ of the projective geometry can be expressed by a pair $(A, \alpha) \in \mathrm{P}\Gamma \mathrm{L}_{n}$. Then
\begin{align*}
d_S( \iota(\mathcal{U}), \iota(\mathcal{V}) ) &= 
d_S( \alpha(\mathcal{U}A), \alpha(\mathcal{V}A) ) \\ 
&= \dim(\alpha(\mathcal{U}A)) + \dim(\alpha(\mathcal{V}A)) - 2\dim(\alpha(\mathcal{U}A) \cap \alpha(\mathcal{V}A) )\\
&= \dim(\mathcal{U}) + \dim(\mathcal{V}) - 2\dim(\alpha((\mathcal{U} \cap \mathcal{V})A))\\ 
&= d_S( \mathcal{U}, \mathcal{V})
\end{align*} 
thus $\iota$ is an isometry with $\iota(\{0\}) = \{0\}$.
\end{enumerate}
\end{proof}

\begin{cor}
 Every isometry $\iota$ on $\ProG$, where $n>2$, with $\dim(\U)=\dim(\iota(\U))$ for any $\U\in \ProG$ is induced by a semilinear transformation $(A,\alpha)\in\mathrm{P}\Gamma \mathrm{L}_{n}$.
\end{cor}

From now on assume that $n>2$. This is no real restriction, because for application, subspace codes in an ambient space of dimension $2$ are not interesting since the only non-trivial subspaces are the one-dimensional ones. In that case neither the transmission rate is improved compared to forwarding, nor is error-correction possible.

\begin{defi}
\begin{enumerate}
\item
Two codes $\C_{1}, \C_{2} \subseteq \ProG$ are \emph{linearly isometric} if there exists $A\in \PGL{n}$ such that $\C_{1}=\C_{2}A$. Since it is the orbit of $\PGL{n}$ on the code, the set of all linearly isometric codes is denoted by $\C_{1} \PGL{n}$.
\item
We call $\C_{1}$ and $\C_{2}$ \emph{semilinearly isometric} if there exists $(A, \alpha) \in \PGammaL{n}$ such that $\C_{1}= \C_{2} (A,\alpha)$. The set of all semilinearly isometric codes is denoted by $\C_{1}\PGammaL{n}$.
\end{enumerate}
\end{defi}

Clearly linear and semilinear isometry are equivalence relations, so it makes sense to speak of classes of (semi-)linearly isometric codes. Note, that the isometries are independent of the underlying metric.

A lattice point-of-view of the isometries of subspace codes can be found in \cite{traut11}.

\begin{defi}
The set 
\[ \SAut(\C) := \Stab{\GammaL{n}}{\C} := \{(A, \alpha) \in \GammaL{n} \mid \C (A, \alpha) =\C\}\]
is a subgroup of $\GammaL{n}$ and is called the \emph{semi-linear automorphism group} of the subspace code $\C$. The \emph{(linear) automorphism group} of $\C$ is defined as
\[\Aut(\C) := \Stab{\GL{n}}{\C} := \{A \in \GL{n} \mid \C A =\C\}\]
and is a subgroup of $\SAut(\C)$.
\end{defi}
Note, that $\Aut(\U G)$  contains every subgroup of $\GL{n}$ that generates the orbit $\U G$.

\begin{lem}
For a subspace code 
$\C :=\{\rs(U_{i}) \mid i=1,\dots, m\} \subseteq \G$ we know that
\[\bigcap_{i=1}^{m}\Stab{\GL{n}}{\rs(U_{i})} \subseteq \Aut(\C).\]
Since
\[B\in \bigcap_{i=1}^{l}\Stab{\GL{n}}{\rs(U_{i})}  \iff \exists A\in\GL{k} : AU_{i} = U_{i} B \;\;\forall i=1,\dots,l \]
we conclude that in particular $\lambda I_{n} \in \Aut(\C)$ 
for all $\lambda \in \F_{q}\setminus\{0\} =: \F_q^{\ast}$.
\end{lem}
Therefore, one can replace the projective groups with $\GL{n}$ and $\GammaL{n}$ when computing isometry classes and automorphism groups of subspace codes.



%

\section{Isometry and Automorphisms of Known Code Constructions}\label{sec:isoex}

In this section we will examine the isometries and automorphism groups of some known classes of constant dimension codes, namely spread codes, orbit codes and lifted rank metric codes.

\subsection{Spread codes}

Spreads of vector spaces are well-known geometrical objects, defined to be partitions of the non-zero elements of a given vector space into subspaces (without the zero-element) of that vector space of a fixed dimension. I.e. a $k$-spread of $\F_{q}^{n}$ is a set of subspaces of dimension $k$ such that they pairwise intersect only trivially and they cover the whole vector space $\F_{q}^{n}$. Thus, a spread exists if and only if $k|n$ and is a subset of $\G$, i.e. it can be used as a constant dimension code. In this case we speak of a \emph{spread code}. A spread code has cardinality $(q^n-1)/(q^k-1)$ and minimum distance $2k$.

Different constructions for these codes are known and have been studied from a coding perspective, e.g. in \cite{go11,ma08p,ma11j}.

The trivial cases are $k=1$ where the spread corresponds to the projective space and $k=n$ where the spread has one element, namely the whole space. 

One way of constructing spreads is the $\F_{q^{k}}$-linear representation of $\F_{q^{n}}$: Since $k|n$ we can consider $\F_{q^{n}}$ as an extension field of $\F_{q^{k}}$ of degree $l:=n/k$, which is isomorphic to the vector space $\F_{q^{k}}^{l}$. In this vector space consider the trivial spread of all one-dimensional subspaces. Each of these lines over $\F_{q^{k}}$ can now be considered as a $k$-dimensional subspace over $\F_{q}$. Since the lines of $\F_{q^{k}}^{l}$ intersect only trivially and with a simple counting argument it follows that the corresponding $k$-dimensional subspaces of $\F_{q}^{n}$ form a spread.

We call a spread code \emph{Desarguesian} if it is an $\F_{q^{k}}$-linear representation of $\F_{q^{n}}$, or if it is a column permutation of such a code. 

\begin{thm}
All Desarguesian spread codes are linearly isometric.
\end{thm}
\begin{proof}
Since there is only one spread of lines in $\F_{q^{k}}^{l}$, different Desarguesian spreads of $\F_{q}^{n}$ can only arise from the different isomorphisms between $\F_{q^{k}}$ and $\F_{q}^{k}$. As the isomorphisms are linear maps, there exists a linear map between the different spreads arising from them.
\end{proof}

In general, not all spreads are linearly isometric but in the special case of $q=2, k=2, n=4$ they actually are, which can be seen as follows. From \cite[Lemma 17.1.3]{hi85} we know that every spread in $\mathcal{G}_q(2,4)$ is regular. Since in $\mathcal{G}_2(k,2k)$ a spread is Desarguesian if and only if it is regular \cite[p. 207]{hi85}, we know that every spread is Desarguesian. Hence all spreads in $\mathcal{G}_q(2,4)$ are linearly isometric.

We will now investigate the automorphism groups of Desarguesian spreads. 

\begin{thm}
The linear automorphism group of a Desarguesian spread code $\C \subseteq \G$ is isomorphic to $\GL{\frac{n}{k}}(q^k) \times Gal(\F_{q^{k}}, \F_q)$.
\end{thm}
\begin{proof}
Let $l:=n/k$.
We want to find all $\F_q$-linear bijections of $\mathbb{P}^{l-1}(\F_{q^k})$. We know that $\PGL{l}(q^k)$ is the groups of all $\F_{q^k}$-linear bijections of $\mathbb{P}^{l-1}(\F_{q^k})$. Thus, $\PGL{l}(q^k) \times Gal(\F_{q^{k}},\F_q)$ is the set of all $\F_q$-linear bijections of $\mathbb{P}^{l-1}(\F_{q^k})$. 
It follows that in the affine space the linear automorphism group of such a spread is isomorphic to $\GL{l}(q^k) \times Gal(\F_{q^{k}}, \F_q)$.
\end{proof}

\begin{cor}
Let $\mathcal{S}$ be a Desarguesian spread code in $\G$. Then 
$$|\Aut(\mathcal{S})|=k\prod_{i=0}^{\frac{n}{k}-1} q^{n}-q^{ki}.$$
\end{cor}
\begin{proof}
Follows from the fact that 
$|Gal(\F_{q^k}, \F_q)|=k$ and $|\GL{\frac{n}{k}}(q^k)|= \prod_{i=0}^{n/k-1} {(q^k)}^{\frac{n}{k}}-{(q^k)}^i$.
\end{proof}

Since $\F_{q^{k}}$ is isomorphic to $\F_{q}[\alpha]$ where $\alpha$ is a root of an irreducible polynomial $p(x)$ of degree $k$ but also to $\F_{q}[P]$ where $P$ the companion matrix of $p(x)$, we get:

\begin{cor}
The automorphism group of a Desarguesian spread code in $\G$ is generated by all elements in $\GL{n}$ where the $k\times k$-blocks are elements of $\F_q[P]$ and block diagonal matrices where the blocks represent an automorphism of $\F_{q^k}$.
\end{cor}

Another point of view of the construction of a Desarguesian spread can be found in \cite{ma08p}, where the generator matrices of the code words are of the type
\[U=\left[\begin{array}{ccccc}
B_{1} & B_{2} & \dots & B_{l}
\end{array}\right]\]
where the blocks $B_{i}$ are an element of $\F_{q}[P]$ and $P$ is a companion matrix of an irreducible polynomial of degree $k$.
To stay inside this structure (i.e. to apply an automorphism) we can permute the blocks, do block-wise multiplications or do block-wise additions with elements from $\F_q[P]$. This coincides with the structure of the automorphism groups from before.

This result is depicted in the following Examples.

\begin{ex}\label{ex1}
Consider $\mathcal{G}_{2}(2,4)$. The only binary irreducible polynomial of degree $2$ is $p(x)=x^{2}+x+1$, i.e. 
\[P=\left( \begin{array}{cc}
0 & 1 \\
1 & 1 
\end{array}\right) . \]
The respective spread code is 
\[\C=\{\rs \left[\begin{array}{ccccc} I & 0
\end{array}\right],
\rs \left[\begin{array}{ccccc} I & I
\end{array}\right],
\rs \left[\begin{array}{ccccc} I & P
\end{array}\right], 
\rs \left[\begin{array}{ccccc} I & P^{2}
\end{array}\right],
\rs \left[\begin{array}{ccccc} 0 & I
\end{array}\right]\}\]
and its automorphism group has $360$ elements:
\[\Aut(\C)=\left\langle
\left( \begin{array}{ccccc}
 & I \\
I &  
\end{array}\right),
\left( \begin{array}{ccccc}
I &  \\
 & P 
\end{array}\right),
\left( \begin{array}{ccccc}
I & P \\
 & I 
\end{array}\right),
\left( \begin{array}{ccccc}
Q &  \\
 & Q 
\end{array}\right)\right\rangle\]
where $Q=\left( \begin{array}{cc}
1 & 0 \\
1 & 1 
\end{array}\right) \in \GL{2}$ represents the only non-trivial automorphism of $\F_{2^2}$, i.e. $x\mapsto x^2$.
\end{ex}

A different approach of finding the automorphism group of a spread in $\mathcal{G}_{2}(2,4)$ can also be found in \cite[Corollary 2]{hi85}.

\begin{ex}\label{ex2}
Consider $\mathcal{G}_{3}(2,4)$ and the irreducible polynomial $p(x)=x^{2}+x+2$, i.e. 
\[P=\left( \begin{array}{cc}
0 & 1 \\
1 & 2 
\end{array}\right)\]
The spread code is 
\[\C=
\rs \left[\begin{array}{ccccc} I & 0
\end{array}\right]\cup
\{\rs \left[\begin{array}{ccccc} I & P^{i}
\end{array}\right] \mid i=0,\dots,7\} \cup
\rs \left[\begin{array}{ccccc} 0 & I
\end{array}\right]\]
and its automorphism group has $11520$ elements:
\[\Aut(\C)=\left\langle
\left( \begin{array}{ccccc}
 & I \\
I &  
\end{array}\right),
\left( \begin{array}{ccccc}
I &  \\
 & P 
\end{array}\right),
\left( \begin{array}{ccccc}
I & P \\
 & I 
\end{array}\right),
\left( \begin{array}{ccccc}
Q &  \\
 & Q 
\end{array}\right)\right\rangle\]
where $Q=\left( \begin{array}{cc}
1 & 0 \\
2 & 2 
\end{array}\right) \in \GL{2}$. Here $Q$ represents the only non-trivial automorphism of $\F_{3^2}$, i.e. $x\mapsto x^3$.
\end{ex}

Note, that in both examples the first element of the generator sets corresponds to swapping the blocks, the second corresponds to multiplication by $P$ and the third element to adding $P$ in the second block of the code word generator matrices.

%
%


\subsection{Orbit codes}

\emph{Orbit codes} are defined as orbits under the group action of the general linear group on the Grassmannian, which is defined as follows:
\begin{align*}
\G \times \GL{n} &\longrightarrow \G \\
(\U , A) &\longmapsto \U A
\end{align*}
They were first defined in \cite{tr10p}.
For more information on orbit codes the reader is referred to \cite{traut11}, where also a similar version of the following fact on isometry of orbit codes can be found:


\begin{thm}
Let $\C_{1}=\U_{1} G$ be an orbit code. Then
$\C_{2}$ is linearly (respectively semilinearly) isometric to $\C_{1}$ if and only if there exists $S\in \GL{n}$ (respectively $S\in \PGL{n}$) such that
\[\C_{2}= \U_{1} S (S^{-1} G S) , \]
i.e. $S^{-1}GS$ is a defining group of $\C_{2}$.
\end{thm}


For a given orbit code $\C\in \G$ we call any subgroup $G\leq \GL{n}$ a \emph{generating group} of $\C$, if $\U G=\C$ for some $\U\in \C$.

If the generating groups of two orbit codes are cyclic, they are conjugate in $\GL{n}$ if and only if the rational canonical forms of their generators have the same number of elementary divisors of the same degree and order and the same respective exponents of the elementary divisors \cite[Section 4]{traut11}.


As mentioned in the beginning, the automorphism groups can be seen as a canonical representative of the generating groups of orbit codes. 
\begin{prop}
\begin{itemize}
\item
Every generating group of an orbit code is a subgroup of the automorphism group. 
\item
Every subgroup of the automorphism group containing a generating group is a generating group. Hence, the automorphism group is a generating group of the orbit code.
\end{itemize}
\end{prop}
\begin{proof}
\begin{itemize}
\item
If $\C =\U G$, then $\C G = \U G G = \U G$.
\item
Let $G$ be a generating group of $\C$ and $G\leq H \leq \Aut(\C)$. Hence, $\C=\U G$ and $\C H = \C$. This implies that $\U H = \U GH = \C H =\C$, since $G$ is a subgroup of $H$.
\end{itemize}
 \end{proof}

The question of finding elements of the automorphism group can be translated into a stabilizer condition of the initial point of the orbit.

\begin{prop}
$A\in \GL{n}$ is in the automorphism group of $\C=\U G $ if and only if for every $B' \in \GL{n}$ there exists a $B''\in \GL{n}$ such that
\[B'AB'' \in \Stab{\GL{n}}{\U}.\]
\end{prop}

\begin{proof}
\begin{align*}
A\in \Aut(\C) &\iff \C A=\C \\
&\iff \forall B'\in G \:\exists B^{*}\in G :\; \U B' A = \U B^{*}\\
&\iff \forall B'\in G \:\exists B^{*}\in G :\; \U B' A {B^{*}}^{-1} =\U
\end{align*}
The statement follows with $B'':={B^{*}}^{-1} \in G$.
\end{proof}

 \subsection{Lifted rank-metric codes}
 
Rank-metric codes are matrix codes, i.e. subsets of $\Mat{k}{m}$ (in this work we will restrict ourselves to the case $k\leq m$) equipped with the rank distance
\[d_{R}(U,V):=\mathrm{rank}(U-V) \quad \textnormal{ for } U,V \in \Mat{k}{m}.\] 
Naturally such a matrix code can also be seen as a block code in $\F_{q^{k}}^m$. We will denote rank-metric codes by $\C_{R}$.



The isometry of rank-metric codes has already been studied in \cite{berger}:

\begin{lem}
\begin{enumerate}
\item
The set of $\F_{q^{k}}$-linear isometries on $\F_{q^{k}}^{m}$ equipped with the rank-metric is $\mathcal{R}^{lin}(\F_{q^{k}}^{m}):= \GL{m}(q) \times \F_{q^{k}}^{*} $.
\item
The set of $\F_{q^{k}}$-semilinear isometries on $\F_{q^{k}}^{m}$ equipped with the rank-metric is $\mathcal{R}^{semi}(\F_{q^{k}}^{m}):=\left(\GL{m}(q) \times \F_{q^{k}}^{*}\right) \rtimes \Aut(\F_{q^{k}}) $.
\end{enumerate}
\end{lem}
For the matrix representation of rank-metric codes we can replace $\F_{q^{k}}$ with $\F_{q}[P]$ where $P$ is the companion matrix of an irreducible polynomial of degree $k$. The multiplication with elements from $\F_{q}[P]$ is done from the left.

Note, that an $\F_{q^{k}}$-linear map is also $\F_{q}$-linear. On the other hand, there might be other $\F_{q}$-(semi-)linear isometries than the ones mentioned before.

One can create constant dimension codes from a given rank-metric code, as explained in the following.

\begin{lem}\cite{si08a}
Let $\C_{R} \subseteq \Mat{k}{n-k}$ be a rank-metric code with minimum distance $d$. Then the lifted code 
\[\C = \{\rs \left[ \:I_{k} \;\; A\: \right] \mid A \in \C_{R}\}\]
is a constant dimension code in $\G$ with minimum distance $2d$.
\end{lem}

\begin{prop}\label{isolift}
The following elements map a lifted rank-metric code to another lifted rank-metric code with the same parameters and are semilinear isometries:
\[\left\{(\left(\begin{array}{cccc}
I_{k} &  \\
      & A 
\end{array}\right), \alpha )
\mid A \in \GL{n-k}, \alpha \in \Aut(\F_{q})\right\}\]
For $\alpha=id$ they are linear isometries.
\end{prop}

\begin{proof}
Follows from the block matrix multiplication rules with
\[\left[\begin{array}{cc}
I_{k} & B
\end{array}\right]
\left(\begin{array}{cccc}
I_{k} &  \\
      & A 
\end{array}\right) =
\left[\begin{array}{cc}
I_{k} & BA
\end{array}\right] \]
and the fact that $A$ is a rank-metric isometry. Moreover, $\alpha I_{k} = I_{k}$ and $\alpha$ is a rank-metric isometry since $\Aut(\F_{q^{k}})\supseteq \Aut(\F_{q})$.
\end{proof}

\begin{cor}
If two rank-metric codes are $\F_{q^k}$-linearly isometric in the rank-metric space, their lifted codes are linearly isometric in the Grassmannian.
\end{cor}
\begin{proof}
Let $\C_{R}$ and $\C'_{R}$ be two linearly isometric rank-metric codes, i.e. $\C'_{R}= P' \C_{R} A$ with $P' \in \F_{q}[P]$ and $A\in \GL{n-k}$. Then the lifted code of $\C'_{R}$ is 
\begin{align*}
\C'&= \{\rs \left[ \begin{array}{cc} I_{k} & R' \end{array}\right] \mid R' \in \C_{R'}\}\\
&=\{\rs \left[ \begin{array}{cc} I_{k} & P'RA \end{array}\right] \mid R \in \C_{R}\}\\
&=\{\rs \left[ \begin{array}{cc} P'^{-1} & R \end{array}\right] \mid R \in \C_{R}\} \left(\begin{array}{cc}  I_{k}&  \\ & A  \end{array}\right)\\
&=\{\rs \left[ \begin{array}{cc} I_{k} & R \end{array}\right] \mid R \in \C_{R}\} \left(\begin{array}{cc}  P'^{-1}&  \\ & A  \end{array}\right)\\
&=\C \left(\begin{array}{cc}  P'^{-1}&  \\ & A  \end{array}\right)
\end{align*} 
where $\C$ is the lifted code of $\C_{R}$. Hence, the lifted codes are linearly isometric.
\end{proof}

Naturally, there are codes that are linearly isometric to a lifted rank-metric code but are not a lifted rank-metric code itself.


We can use the knowledge of the automorphism group of a rank-metric code also for finding the automorphism group of the respective lifted rank-metric code. For this denote by $\Aut_{R}$ the automorphism group of the rank-metric code.

\begin{prop}\label{lift1}
Let $\C_{R} \subseteq \Mat{k}{(n-k)}$ be a rank-metric code and $\C$ its lifted code. Then
\[\left\{ \left( \begin{array}{cc}
I_{k} &  \\
 & R 
\end{array}\right) \mid R \in \Aut_{R} (\C_{R})\right\} \subseteq \Aut(\C) .\]
\end{prop}

\begin{proof}
It holds that
\[\{\left[\begin{array}{cc}
I_{k} & B
\end{array}\right] \mid B\in \C_{R}\}
\left(\begin{array}{cccc}
I_{k} &  \\
     & R 
\end{array}\right) =
\{\left[\begin{array}{cc}
I_{k} & BR
\end{array}\right]\mid B\in \C_{R}\} .\]
Since $R\in \Aut_{R}(\C_{R})$, this set is equal to the original one.
\end{proof}

\begin{thm}\label{lift2}
Let $\C_{R} \subseteq \Mat{k}{(n-k)}$ be a rank-metric code and $\C$ its lifted code. Then
\[\left\{ \left( \begin{array}{cc}
I_{k} &  \\
 & A 
\end{array}\right) \mid A \in \GL{n-k}\right\} \cap \Aut(\C) = \left\{ \left( \begin{array}{cc}
I_{k} &  \\
 & R 
\end{array}\right) \mid R \in \Aut_{R}(\C_{R}) \right\}.\]
\end{thm}
\begin{proof}
From Proposition \ref{lift1} we know that the right side is included in the left. Furthermore, 
\[\rs\left[\begin{array}{cc}I_{k} &B_{1}\end{array}\right]  \left( \begin{array}{cc}
I_{k} &  \\
 & A
\end{array}\right) = \rs\left[\begin{array}{cc} I_{k} & B_{2}\end{array}\right]  \]
\[\iff  \exists C_{1},C_{2} \in \GL{k} : \left[\begin{array}{cc} C_{1}&  C_{1}B_{1}\end{array}\right]  \left( \begin{array}{cc}
I_{k} &  \\
 & A
\end{array}\right) =\left[\begin{array}{cc} C_{2} &C_{2}B_{2}\end{array}\right]  \]
\[\iff  C_{1}=C_{2} \quad\wedge\quad B_{1}A = B_{2}
\]
i.e. if $\left( \begin{array}{cc}
I_{k} &  \\
 & A 
\end{array}\right)\in \Aut(\C) $, then $A\in \Aut_{R}(\C_{R})$.
\end{proof}
Hence, if we know the automorphism group of a lifted rank-metric code, we also know the automorphism group of the rank-metric code itself.

\begin{ex}
Consider the rank-metric code 
\[\C_R = \left\{\left( \begin{array}{cc}
1 & 0  \\
0 & 1 
\end{array}\right),
\left( \begin{array}{cc}
1 & 1  \\
0 & 1 
\end{array}\right),
\left( \begin{array}{cc}
0 & 1  \\
0 & 1 
\end{array}\right),
\left( \begin{array}{cc}
0 & 0  \\
0 & 1 
\end{array}\right)\right\}\]
with four elements and minimum rank distance $1$ over $\F_2$. Its automorphism group is
\[\Aut_R(\C_R) = \left\{\left( \begin{array}{cc}
1 & b  \\
0 & 1 
\end{array}\right)\mid b \in\F_2 \right\}  .\] 
Let $\C$ be the lifted code of $\C_R$ in $\mathcal{G}_2(2,4)$. Then
\[\Aut(\C) = \left\langle \left( \begin{array}{cccc}
1 & 0 & 0 & 0  \\
0 & 1 & 0 & 0 \\
0 & 1 & 1 & 0 \\
0 & 0 & 0 & 1 
\end{array}\right),
\left( \begin{array}{cccc}
1 & 0 & 0 & 0  \\
0 & 1 & 0 & 0 \\
0 & 0 & 1 & 1 \\
0 & 0 & 0 & 1 
\end{array}\right), \right.\]\[\left.
\left( \begin{array}{cccc}
1 & 0 & 0 & 0  \\
0 & 0 & 1 & 0 \\
0 & 1 & 0 & 0 \\
0 & 1 & 1 & 1 
\end{array}\right),
\left( \begin{array}{cccc}
1 & 1 & 0 & 0  \\
0 & 1 & 0 & 0 \\
0 & 1 & 1 & 0 \\
0 & 0 & 0 & 1 
\end{array}\right)
 \right\rangle\]
with $|\Aut(\C)|=192$. The second generator and the identity matrix are the corresponding elements described in Theorem \ref{lift2}.
\end{ex}



\section{Conclusion}\label{sec:conclusion}

In this work we investigated linear and semilinear isometry, as well as linear and semilinear automorphism groups, for general network codes, i.e. sets of vector spaces over a finite field.  We showed that the subset-relation-and-dimension-preserving isometries correspond exactly to the general (semi-) linear group.

In Section \ref{sec:isoex} we showed some theoretical results and examples of isometry classes and automorphism groups of some known constructions of constant dimension codes, namely spread codes, orbit codes and lifted rank metric codes.

The isometry classes indicate how many non-equivalent different codes for given size and minimum distance can be found. On the other hand, the automorphism groups are useful for counting how many different codes there are in the same isometry class of a given code.

Moreover, the automorphism groups of orbit codes function as a canonical generating group to compare orbit codes among each other. 

More research can be done in finding theoretical results on the automorphism groups of constant dimension codes. E.g. one could study the automorphism groups of non-Desarguesian spreads or try to find a family of orbit codes that have a certain automorphism group. Moreover, one could study the isometry and automorphism groups of non-constant dimension codes. 

In general, it might not always be possible to compute the automorphism group of an arbitrary subspace code or check if two given codes are isometric. In this case, the algorithm of T. Feulner in \cite{thomas} can be used to do these computations.

For future work it would be interesting to see how the knowledge of the automorphism group of a given constant dimension code (or a general subspace code) can be helpful for decoding, as it is in the classical block code case.

\section*{Acknowledgements}
The author thanks Thomas Feulner, without whom this work would not have been possible, as well as Joachim Rosenthal and Leo Storme for their advice.

\bibliographystyle{plain}
\bibliography{our}

\end{document}